\documentclass[fleqn,print]{aa}  

\usepackage{graphicx,color}
\usepackage[varg]{txfonts}
\usepackage{natbib,xspace}
\usepackage{stmaryrd}
\usepackage{siunitx}
\usepackage[switch]{lineno}
\usepackage{rotating}
\usepackage{dcolumn}
\usepackage{upgreek}
\usepackage{hyperref}
\usepackage{wasysym}
\usepackage{multirow}
\newcommand{\mr}[1]{\multirow{2}{*}{#1}}
\usepackage[ulem=normalem]{changes}

\newcolumntype{.}{D{.}{.}{-1}}

\newcommand*\mean[1]{\overline{#1}}

\providecommand\aap{Astron. \& Astrophys.}

\providecommand\aj{Astron. J.}
\providecommand\apj{Astrophys. J. }
\providecommand\apjl{Astrophys. J. Lett.}
\providecommand\apjs{Astrophys. J. Supp.}

\providecommand\asr{Adv. Space Res.}

\providecommand\icarus{Icarus}%

\providecommand\mnras{Mon. Not. R. Astron. Soc. }
\providecommand\nat{Nature}

\providecommand\planss{Planet. Space Sci.}
\providecommand\prd{Phys.~Rev.~D}%
\providecommand\ssr{Space~Sci.~Rev.}%

\begin{document}

\title{Modeling the astrosphere of LHS~1140}
\subtitle{On the differences of 3D (M)HD single- and multifluid simulations and the consequences for exoplanetary habitability}
\author{K. Scherer \inst{1,2,3} \and K. Herbst \inst{4,5} \and
    N.E. Engelbrecht\inst{6} \and S.E.S. Ferreira\inst{6} \and J. Kleimann \inst{1,3} \and J. Light\inst{6}}
\institute{Institut f\"ur Theoretische Physik IV, Ruhr-Universit\"at Bochum, 44780 Bochum, Germany,
\email{kls@tp4.rub.de} 
\and
 Research Department ``Plasmas with Complex Interactions,'' Ruhr-Universit\"at Bochum, Germany 
\and
Ruhr Astroparticle and Plasma Physics Center (RAPP Center), 44780 Bochum, Germany
\and
Institut f\"ur Experimentelle and Angewandte Physik, Christian-Albrechts-Universit\"at zu Kiel, Germany
\and
Institut f\"ur Planetenforschung (PF). Deutsches Zentrum f\"ur Luft- und Raumfahrt (DLR). Rutherfordstr. 2, 12489 Berlin. Germany
\and
Centre for Space Research, North-West University, 2520, Potchefstroom, South Africa}

\date{Received: accepted}

\abstract
{The cosmic ray (CR) flux, as well as the hydrogen flux into the atmosphere of an exoplanet, can change the composition of the atmosphere. Here, we present the CR and hydrogen flux on top of the atmosphere. To do so, we have to study the 3D multifluid MHD structure of astrospheres.}
{We discuss the shock structure of the stellar wind of LHS~1140 using four different models: hydrodynamic and ideal magnetohydrodynamic single-fluid models, as well as multifluid models for both cases, including a neutral hydrogen flow from the interstellar medium. The CR flux in a multifluid model as well as the ionization rate in an exoplanetary atmosphere are also presented.  }
{The astrosphere is modeled using the 3D Cronos code, while the CR flux at LHS~1140~b is calculated using both a 1D and a 3D stochastic galactic CR modulation code. Finally, the atmospheric ionization and radiation dose is estimated using the AtRIS code. }
{It is shown that the 3D multifluid positions of the termination shock differ remarkably from those found in the 3D ideal-single fluid hydrodynamic case. CR fluxes computed using a 1D approach are completely different from those calculated using the 3D modulation code and show an essentially unmodulated spectrum at the exoplanet in question. Utilizing these spectra, ionization rates and radiation exposure within the atmosphere of LHS~1140~b are derived.}
{It is shown that the multifluid MHD termination shock distances differ remarkably from those of other models, especially from an analytic approach based on ideal single-fluid hydrodynamics. The termination shock, astropause, and bow shock distances must be taken from the 3D multifluid MHD model to determine the CR fluxes correctly. Moreover, because of the tiny astrosphere, the exoplanet is submerged in the neutral hydrogen flow of the interstellar medium, which will influence the exoplanetary atmosphere. A 3D approach to Galactic cosmic ray (GCR) modulation in astrospheres is also necessary to avoid unrealistic estimates of GCR intensities. Since atmospheric chemistry processes, and with that, the derivation of transmission spectra features and biosignature information, strongly depend on atmospheric ionization, our results highlight that reliable GCR-induced background radiation information is mandatory, particularly for inactive cool stars such as LHS~1140.}
\maketitle

\section{Introduction}
LHS~1140, a 5\,Gyrs old \citep{Dittmann-etal-2017} M4.5-class star\footnote{\href{http://simbad.u-strasbg.fr/simbad/}{http://simbad.u-strasbg.fr/simbad/}}, is located at a distance of $(12.47 \pm 0.42)$\,pc. As of today, 
three planets have been confirmed to orbit LHS~1140: LHS~1140~b with a rotation period of 24.7 days, LHS~1140~c orbiting within 3.77 days, and LHS~1140~d with a rotation period of 78.9 days \citep{Lillo-BoxEA2020}. With the help of ESPRESSO, TESS, and HARPS observations, \citet{Lillo-BoxEA2020} determined their masses within 9\% precision. They found the masses to be $(6.48 \pm 0.46)\, M_{\oplus}$ (LHS\,1140 b), $(1.78 \pm 0.17)\, M_{\oplus}$ (LHS~1140~c), and $(4.8 \pm 1.1)\, M_{\oplus}$ (LHS~1140~d). Thus, while LHS~1140~b might be a temperate mini-Neptune or a water world \citep{CadieuxEA2023}, LHS~1140~c might be an Earth twin \citep{Lillo-BoxEA2020}.  However, although only LHS~1140~b, with a distance of $(0.0875 \pm 0.0041)$\,au, is well within the conservative habitable zone of the stellar system \citep{HillEA2023}, \citet{Lillo-BoxEA2020} found that the water content in both LHS~1140~b and c is compatible with a deep ocean layer.

It has been shown, both by measurements and modeling efforts, that solar and Galactic cosmic rays (GCRs) have a clear impact on Solar system planets like Earth \citep[e.g.,][]{Banjac-Herbst-Heber-2019}, Venus \citep[ e.g.,][]{Nordheim-etal-2015, Herbst-etal-2019b, HerbstEA2020}, and Mars \citep[e.g.,][]{Guo-etal-2019}. Thereby, incoming charged high-energy particles induce secondary particle cascades, which, in turn, can cause the ionization of the planetary atmosphere. This further leads to drastic changes in the atmospheric evolution, climate, and photochemistry, and thus, in turn, in the atmospheric biosignatures \citep[e.g.,][]{Scheucher-etal-2018, grenfell2019exoplanetary, Herbst-etal-2024}, as well as the altitude-dependent atmospheric radiation dose \citep[see, e.g.,][]{Atri-2020, Herbst-etal-2019b, HerbstEA2020}. This, however, strongly depends on solar activity and, with that, on the size of the heliosphere, which shields the solar system planets from the incoming GCRs. This heliospheric shielding is important when it comes to the habitability within our Solar system and must be considered when the habitability of exoplanets is discussed \citep{EngelbrechtEA24}. Therefore, studying the astrospheres of cool stars is of utmost importance.

The astrosphere of LHS~1140 has hitherto received relatively little attention in the literature. The single-fluid magnetohydrodynamic (MHD) simulations of \citet{HerbstEA20} revealed an extraordinarily small astrosphere.  
However, the multifluid 3D MHD modeling of an astrosphere is essential
because both the stellar wind and the interstellar medium (ISM) carry
along a magnetic field. On top of that, the distances of the shock
structures will differ remarkably from those calculated using a  hydrodynamic (HD) approach. 
Moreover, the hydrogen flux can reach deeply into the astrosphere, and in the case of LHS~1140~b, the exo-atmosphere is flooded with the interstellar (atomic) hydrogen flow  (see below). The first results based on single-fluid 3D MHD have been presented by \citet{HerbstEA2020}. This study shows that these results differ significantly from the multifluid 3D MHD results. For the advanced modeling of the heliosphere (the astrosphere around the Sun), see the recent review by \citet{Richardson-etal-2023}.

A growing number of studies are investigating the potential influence of GCRs on the habitability of Earth-like exoplanets in a variety of astrospheres \citep{HerbstEA22}. GCRs undergo various processes when they enter an astrosphere. These include diffusion due to their being scattered by turbulent irregularities in the astrospheric magnetic field \cite[e.g.,][]{Shalchibook}; drifts due to gradients in said field as well as due to the potential presence of neutral sheet structures \cite[e.g.,][]{EngelbrechtEA19}; adiabatic energy changes and convection with the stellar wind \cite[e.g.][]{JokipiiParker70}, amongst others \cite[see][]{Schlickeiser02}. These various processes reduce GCR intensities to varying degrees within the astrosphere in question, a process called modulation which has been extensively studied in the heliospheric context, using numerical models of increasing complexity from the earliest 1D solvers of the \citet{Parker-1965} GCR transport equation to more recent 3D, time-dependent solvers \cite[see, e.g.,][and references therein]{Quenby84,frank,JK00,Kota13,EngelbrechtEA22}.

Prior astrospheric GCR modulation studies, however, have primarily been performed with simple 1D solvers of the \citet{Parker-1965} GCR transport equation, often reporting negligibly small GCR intensities (relative to those observed at $1$~au in the heliosphere), and concluding that any influence these particles could have on exoplanet habitability would probably be negligible \cite[see, e.g.,][]{strum,RL21a,RodgersEA21,MesquitaEA21,MesquitaEA22}. It should be noted, however, that 3D modeling of GCR transport, which can incorporate more transport mechanisms such as GCR drifts and differences in particle diffusion parallel and perpendicular to the astrospheric magnetic field known to play a significant role in the heliospheric transport of these particles \cite[see, e.g.,][and references therein]{EngelbrechtEA22}, has been shown to lead to GCR intensities significantly larger than previously expected. This implies that these particles cannot always be neglected in the context of exoplanetary habitability \citep{EngelbrechtEA24}. Thus far, only one other GCR modulation study has been made for LHS~1140, namely the 1D modulation study of \citet{HerbstEA20}, which reported a very moderate level of modulation at LHS~1140~b, with GCR intensities larger than those typically expected at Earth. The present study considers for the first time modulation in the astrosphere of LHS~1140 in 3D, using the modulation approach  towards solving the Parker transport equation introduced and discussed in detail for Proxima Centauri by \citet{EngelbrechtEA24}. Both single- and multifluid MHD results are employed to obtain inputs for large-scale plasma parameters. These results are then compared with those computed, using the same MHD inputs, with the 1D GCR modulation code of \citet{LightEA22}, to ascertain discrepancies between results calculated with these approaches.

Over the past years, the impact of charged particles on the atmospheres of exoplanets has also attracted growing attention in the exoplanetary community \citep[e.g.,][]{Tabataba-Vakili-etal-2016, Tilley-etal-2017, Alvarado-Gomez-2019, HerbstEA20}. Most recent studies suggest that energetic particles of Galactic and stellar origin may not only strongly affect the exoplanetary atmospheric dynamics, chemistry (and with that biosignatures), and climate \citep[e.g.,][]{Airapetian-etal-2020, Herbst-etal-2019b, Scheucher-etal-2020b, ChenEA21} but also the atmospheric secondary particle environment and the planetary radiation exposure \citep[e.g.,][]{Herbst-etal-2019b, Scheucher-etal-2020a, Herbst-etal-2024}. Thus, answering whether or not a planet is habitable based on the climate, chemistry, and biosignature signals detected with JWST and future missions alone is rather complex, and an interdisciplinary approach incorporating knowledge from the stellar surroundings (i.e., stellar astrospheres) to the exoplanet is mandatory.
 
\section{Hydrodynamic versus magnetohydrodynamic modeling}
\label{sec:2}

Almost all stars produce outflows, which become superfast\footnote{Note that we use the terminology super/subfast for super/subsonic and super/subfast-magnetosonic. This way, the discussion holds true for both HD and MHD.} winds beyond a critical surface. When interacting with the ISM, these winds have to transition from super- to subfast speeds. The transition happens at the termination shock (TS), which in (M)HD is an infinitesimally thin ISM flow is superfast, a bow shock (BS) arises, the superfast to subfast transition of the inflowing ISM. The astropause (AP) is the tangential discontinuity
that separates the interstellar from the stellar environment. There is no mass flux or magnetic flux passing through the AP: The velocity and magnetic field on both sides of the AP are parallel to the AP, but the respective values can differ. The region between the TS and the AP is called the inner astrosheath, where a lot of GCR modulation happens in the heliospheric case. The region between the AP and the BS is called the outer astrosheath. In the inflow region of the outer astrosheath, a hydrogen wall will be created (if the region is large enough). In the tail direction, the TS is replaced by a Mach disk because, for entropy reasons, a standard shock transition is impossible. At the Mach disk arises a so-called triple point from which another tangential discontinuity emerges. This is well-known for the HD case \citep[e.g.,][]{Courant-Friedrichs-1948, Scherer-etal-2020}. To our knowledge, there is no such detailed analysis of the MHD case, but in all our simulations, we see features similar to those in HD. Also, in \citet{Scherer-etal-2020}, the Rankine-Hugoniot relations for the MHD case can be found. For more details on MHD shocks, we refer the reader to the textbook by, e.g., \citet{Goedbloed-etal-2010}.

The general set of the Euler equations, which include the continuity, momentum, energy equation, and, in addition, the induction equation for the magnetic field, are given by

\begin{equation}
\label{eq:mhd}
  \frac{\partial}{\partial t}
  \begin{bmatrix} \rho \\ \rho \, \vec{u} \\ e \\ \vec{B}  \end{bmatrix}
  + \nabla \cdot \begin{bmatrix} 
  \rho \vec{u} \\ 
        \rho \, \vec{u}\otimes\vec{u} + \left(P + \dfrac{B^2}{8\pi}\right) \hat{\vec{I}} -
             \dfrac{\vec{B}\otimes\vec{B}}{4 \pi} \\
        (e + p)\vec{u} - \dfrac{\vec{B}(
                  \vec{B}\cdot\vec{u})}{4 \pi} \\
         \vec{u}\otimes\vec{B} - \vec{B}\otimes\vec{u}
  \end{bmatrix}
 = \begin{bmatrix}
      S^\mathrm{c} \\ \vec{S}^\mathrm{m} \\
      S^\mathrm{e} \\ \vec{0} 
 \end{bmatrix}
\end{equation}
where $\rho$, $\vec{u}$, $P$, and $\vec{B}$ are the mass density, the velocity, the thermal pressure, and the magnetic field vector, and
\begin{equation}
  e = \frac{\rho u^2}{2} + \frac{P}{\gamma-1} + \frac{B^2}{8\pi}
\end{equation}
is the internal energy for an adiabatic index $\gamma$, which we take as $5/3$. The terms on the right-hand side of Eq.~(\ref{eq:mhd}) describe the source terms for charge exchange, electron impact, and photonionization \citep[for details see][]{Scherer-etal-2014}. 

The above set of equations describes the MHD (ionized) fluid while setting $\vec{B}=\vec{0}$ describes the neutral hydrogen fluid\footnote{ When the word "hydrogen" is used, we refer to neutral hydrogen atoms, while "proton" stands for ionized hydrogen.}, and both fluids are coupled via the right-hand side terms. This set of equations is integrated in time on a fixed 3D grid with the Cronos code \citep{Kissmann-etal-2018} until a (sufficiently) steady state has been reached. 
The Cronos code was successfully applied in a series of publications pertaining to the context of the inner \citep[e.g.][]{Czechowski_Kleimann-2017, Wiengarten-etal-2014, Wiengarten-etal-2015} and outer \citet{Kleimann-etal-2023} heliosphere, as well as to astrospheres \citep{HerbstEA20,Scherer-etal-2020,Baalmann-etal-2022}.

To show the difference between the cases with and without a magnetic field as well as those with and without neutral gas, we modeled all four cases. We denote a single-fluid simulation with ``S'' and a multifluid one with ``M.'' These letters are combined with ``H'' for a hydrodynamic (HD) and ``M'' for an MHD simulation, thus we have the descriptive combinations SH, MH, SM, and MM.

The astrosphere is determined by the stellar wind and ISM parameters, especially the ram pressure $P_\mathrm{ram}=\rho u^2/2$. Because the stellar wind is superfast, the TS will be created in front of the inflowing\footnote{The inflow is always parallel to the $x$-axis and flows from the positive (upwind) to the negative $x$ (downwind) axis.} interstellar wind, where the superfast wind becomes subfast. 
A pressure equilibrium surface or tangential discontinuity is generated between the shocked stellar wind and the (shocked) ISM, the astropause (AP). Note that this is not a contact discontinuity. If the inflowing ISM is also superfast another shock is generated, the bow shock (BS). If the fast-magnetosonic Mach number of the ISM is around unity, the BS becomes a bow wave, or if it is subfast, no BS will appear \citep{Fraternale-etal-2023,Fraternale-etal-2023b}. 

\subsection{Single-fluid approach}
The single-fluid approach can be modeled in spherical coordinates. This has the advantage that the inner boundary of the integration area can be chosen to be one-third\footnote{ This factor is based on experience to ensure that the TS is beyond the inner boundary}. In rare cases, a smaller value must be chosen. of the analytic TS distance $r_\mathrm{TS}$ given by \citep{Parker-1963,Wilkin-1996}:
\begin{equation}
  \label{eq:ts_dist}
  r_\mathrm{TS} =
  r_{0} \sqrt{\frac{\rho_{0, \mathrm{sw}}\,u_\mathrm{sw}^{2}}{\rho_\mathrm{ism} \, u_\mathrm{ism}^{2}}} =
  \sqrt{\frac{\dot{M}_\star \, u_\mathrm{sw}}{4\pi \, \rho_\mathrm{ism} \, u_\mathrm{ism}^{2} }}
\end{equation}
with $\rho_{0, \mathrm{sw}}$ the stellar wind density at a reference distance $r_0$, $\rho_\mathrm{ism}$ and $v_\mathrm{ism}$ the constant density and velocity of the ISM, and $\dot{M}_\star$ the stellar mass-loss rate. 
The above TS distance is the only distance that can be calculated for a single-fluid HD model \citep{Scherer-etal-2020}. For all other models (MH, SM, MM), this is not possible (see the discussion below). Equation~(\ref{eq:ts_dist}) is useful to determine the outer integration area.

For the comparison of the models we have used $v_\mathrm{sw}=250\,\si{km/s}$ and $\dot{M}_{\star}=5\cdot 10^{-17} M_\odot/\si{yr}$. These values are slightly lower than those given recently in literature (see Appendix~\ref{app:params}), but for the purposes of comparison, any set of parameters will do. More realistic values are presented in Section~\ref{sec:discussion}. For the more realistic values derived using the approach by \citet{Modi-etal-2023}, $v_\mathrm{sw}=430\,\si{km/s}$ and $\dot{M}_{\star}=2\cdot 10^{-16} M_\odot/\si{yr}$, TS, AP, and BS distances increase by a factor of 3-4. Nevertheless, the LHS\,1140 astrosphere remains so small that for those values neither the GCR fluxes nor the atmospheric ionization rate changes (see below for further discussion).
 
The outer boundary is usually set to 4-5 times the value of $r_\mathrm{TS}$ given by Equation~(\ref{eq:ts_dist}) because the shock structures can reach deep into the ISM during its evolution. Care must be taken when the ISM inflow is subfast (either subsonic in HD or sub-fast-magnetosonic in MHD) because the ISM can be influenced on much larger scales.
Usually a spherical model can run with $1024\times64\times128$ cells. However, to facilitate comparison of results from single-fluid and multifluid models, we have chosen to use a Cartesian grid throughout this paper(see below). 
\begin{figure*}[t!]
    \includegraphics[width=0.95\textwidth]{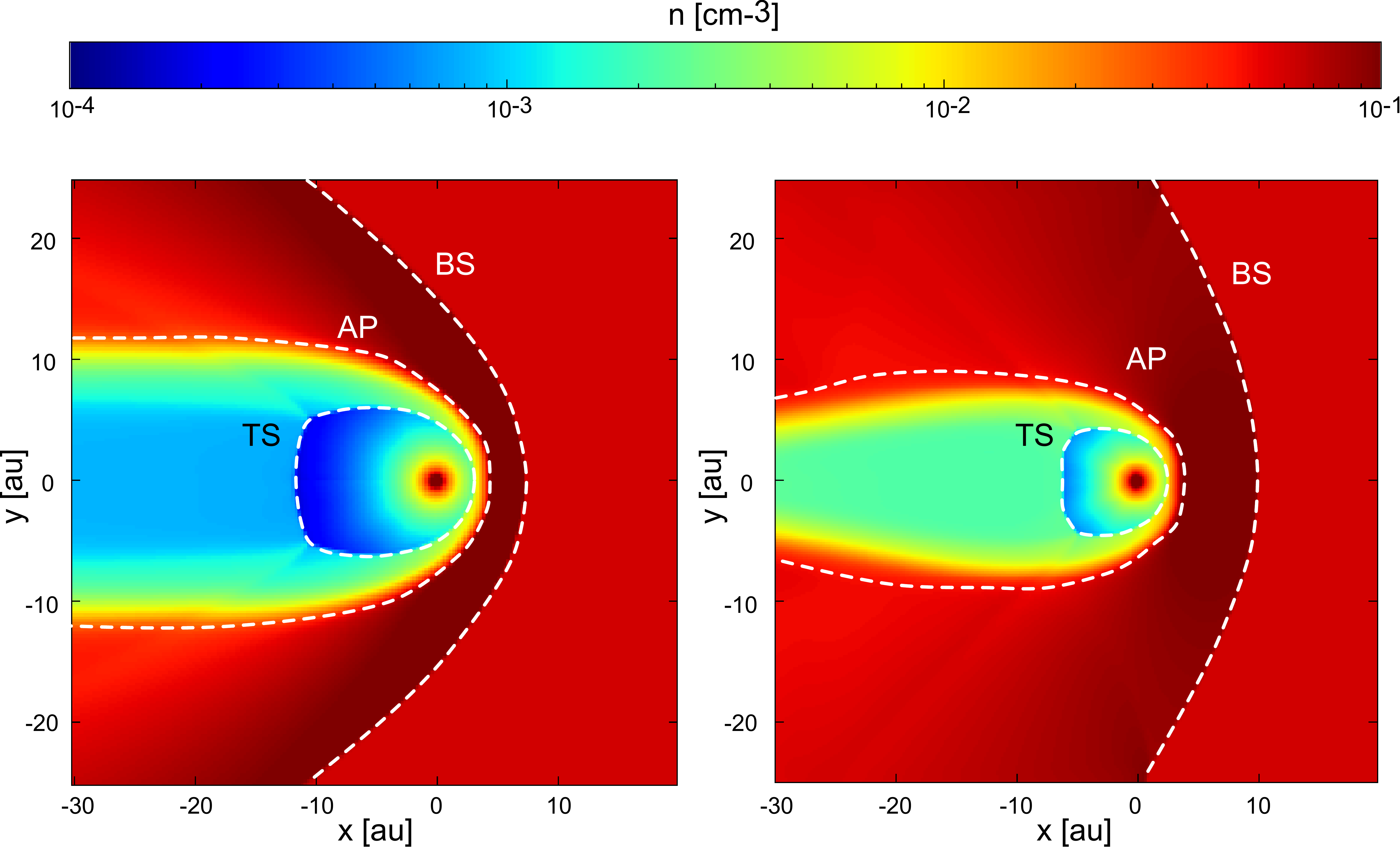}
    \caption{Proton number density for HD runs. Left: single-fluid results (model SH). Right: multifluid results (model MH).
    \label{fig:HD}}
\end{figure*}

\begin{figure*}[t!]
\includegraphics[width=0.95\textwidth]{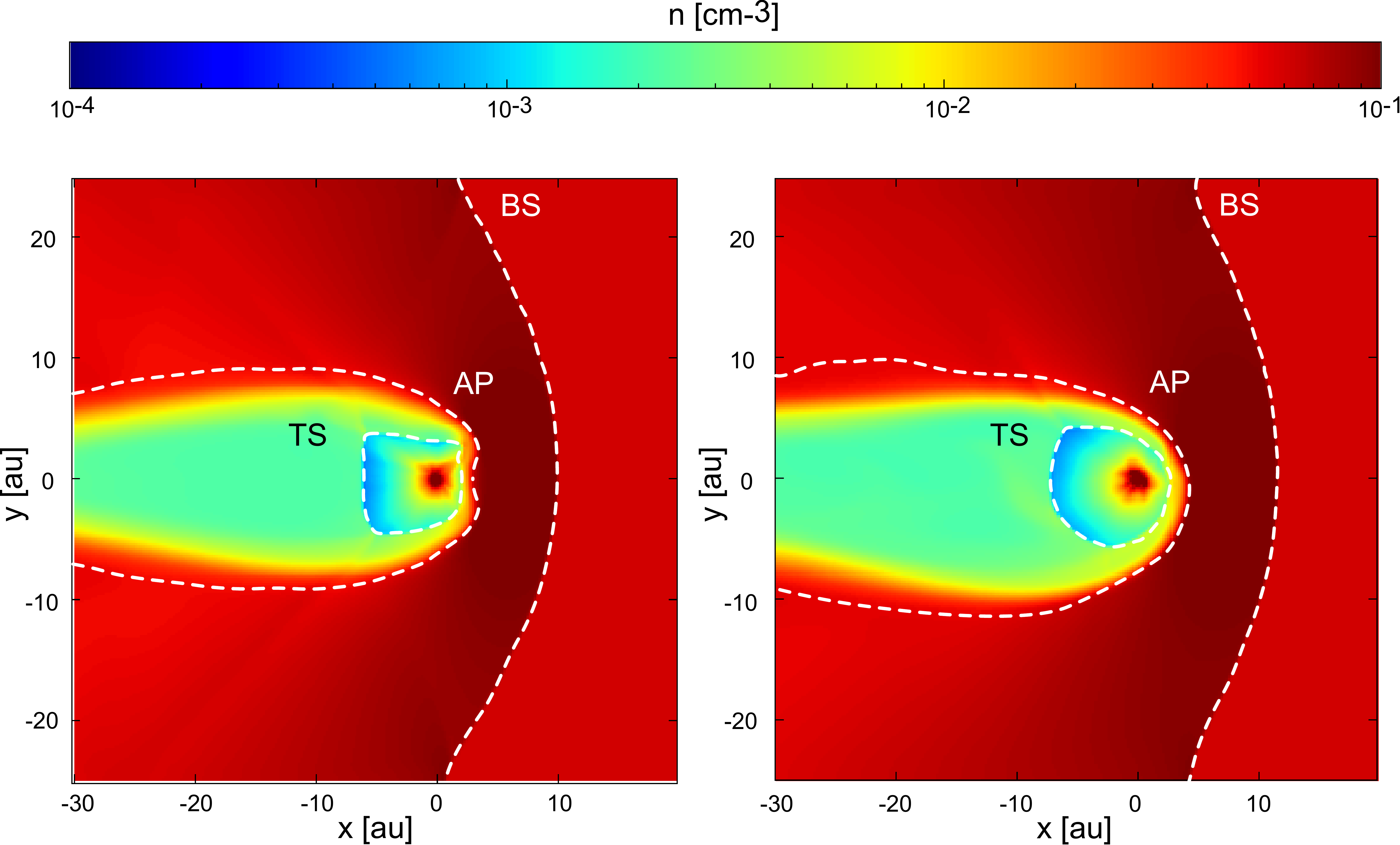}
    \caption{Proton number density for MHD runs. Left: single-fluid results (model SM). Right: multifluid results (model MM).
    \label{fig:MHDp}}
\end{figure*}
\begin{figure*}[!t]
    \centering
    \includegraphics[width=\textwidth]{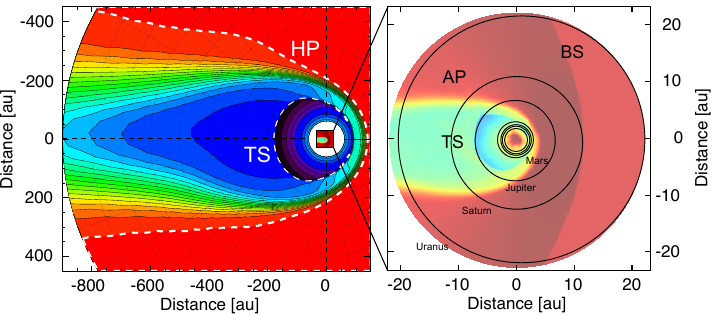}
    \caption{The astrosphere of LHS~1140 compared to the heliosphere \citep[left; e.g.,][]{HerbstEA20} with a zoom-in (right) showing the astrosphere of LHS~1140 in comparison to the Solar system. Shown is the number density in units of cm$^{-3}$. Colors are not to scale.
    \label{fig:1}}
\end{figure*}
\begin{figure}[t!]
    \centering
    \includegraphics[width=\columnwidth]{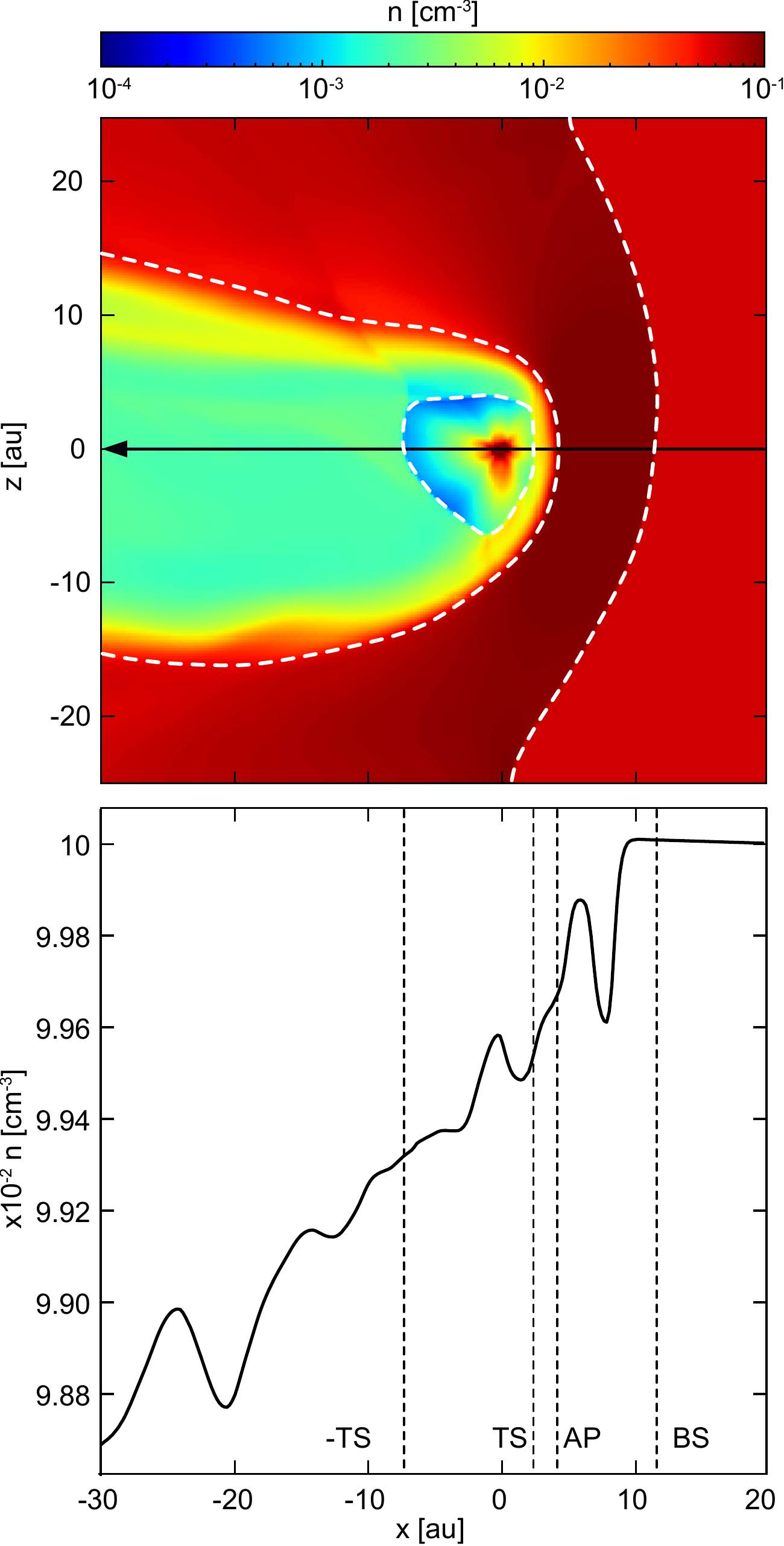}
    \caption{Multifluid runs. Upper panel: The $xz$-plane showing the asymmetry in the tail (with the hydrogen density increased to 1\,\si{cm^{-3}}). Lower panel: The hydrogen density along the $x$-axis. The termination shock (TS), astropause (AP), the bow shock (BS), and the termination shock in the tail direction (-TS) are indicated by the dotted lines. 
    \label{fig:MHDH}}
\end{figure}

\subsection{Multifluid approach}

The multifluid approaches must be simulated on a Cartesian grid because of the required parallelization. The reason is that for  the neutral gas flow, the inner cells in the upwind direction have to be connected to the downwind cells, which is not easy in a spherical grid because it is not known which computational sub-volumes are connected by a neutral streamline. Moreover, when cutting out a sphere, it is unclear how to model the neutral gas inside the inner boundary. Therefore, a Cartesian grid, which lacks an inner (stellar) boundary and thus allows for the neutral component to be seamlessly followed also near the origin, is the best and easiest solution in this case. 

One disadvantage of using the Cartesian grid for large TS, AP, and BS distances is that the number of cells increases dramatically. For the case of LHS~1140 that is fortunately not a problem because of the small extent of its astrosphere, and we use $200^3$ cells, which is less than the number of cells in a standard spherical run. The advantage of using equally spaced Cartesian cells is that the time step does not change when varying the boundaries, while for spherical cells, the time step depends on the smallest cell. Changing the inner boundary changes the time step and the overall run time. Thus, a spherical single-fluid MHD run with a distant inner boundary is computationally much less costly than a Cartesian run. On the other hand, the flux of neutrals requires a Cartesian setup. In the case discussed here, the neutral flux at the position of the exoplanet is roughly the same as the interstellar flux (for further discussion see below).  

For the heliosphere, there are many advanced multifluid models in the literature \citep[see the review by][and references therein]{Richardson-etal-2023}. 
The parameters for the simulations, which are slightly different (the ISM speed) from those used in \citet{HerbstEA20}, are summarized in Appendix~\ref{app:params}.

\section{HD vs. MHD results}
The results of our simulations are shown in Figures~\ref{fig:HD} to~\ref{fig:MHDH}. In the left panels of Figure~\ref{fig:HD} and Figure~\ref{fig:MHDp}, the single fluid- and in the right panels, the multifluid proton number densities are shown. In Figure~\ref{fig:MHDH}, the number density of the neutral hydrogen along the $x$-axis is shown. In the left panel of Figure~\ref{fig:HD}, the Mach disk (and the reflected shock) can be seen, as described in \citet{Scherer-etal-2020}, while in the right panel of Figure~\ref{fig:HD} 
the Mach disk almost disappears. In the multifluid HD model, the TS distances change (see Table~\ref{tab:dist}), especially in the tail direction. Compared to the single-fluid HD model, the BS as well as the astrotail widens, as does the distance between the AP (the yellow boundary) and the BS (dark red boundary). The BS (the dark reddish boundary) in the inflow direction shrinks, as does the entire shock structure. In Table~\ref{tab:dist}, we listed the shortest distances to the TS, AP, and BS. All these distances are measured along the inflow line. This is completely different for the MHD case because now the astrosphere can become asymmetric, and hence, the  TS, AP, and BS distance perpendicular to the inflow axis can be smaller than those on the x-axis. As shown in Table~\ref{tab:dist} these distances differ and cannot be reproduced by Equation~(\ref{eq:ts_dist}). For example, for the heliosphere the TS in an ideal HD model is located at 150\,\si{au}, which agrees nicely with Eq.~(\ref{eq:ts_dist}), but in a multifluid MHD model it is about 75\,\si{au}, which is what is confirmed by Voyager observations \citep[see][and references therein]{Richardson-etal-2023}.

As shown in the left panel of Fig.~\ref{fig:1}, the 3D multifluid MHD LHS~1140 astrosphere is much smaller than the heliosphere. A zoom-in (shown in the right panel) reveals a TS distance of 2.2\,au, an AP distance of 3.6\,au, and a BS at 9.9\,au. This would place the entire astrosphere within the orbit of Uranus.

\begin{table}[t!]
\caption{ \label{tab:dist}TS, AP and BS distances}
    \centering
\begin{tabular}{lrrrc}
    Model           &TS      &AP   &BS  & analytic TS\\
    \hline
    SH   &3.0   &5.1 & 7.5 & 3.3\\
    MH   &2.2   &4.2 & 5.0 & 3.3\\
    SM   &2.4   &4.0 &10.4 & 3.3\\
    MM   &2.2   &3.7 &11.4 & 3.3\\
    MM1  &2.2   &3.7 &10.6& 3.3\\

\hline
     \end{tabular}
    \tablefoot{The shortest distance to the TS, the AP and the BS. All values in units of \si{au}.}
\end{table}

Figure~\ref{fig:MHDH} shows that the hydrogen flux is almost unaffected. The variations along the $x$-axis can be neglected, but a slight increase occurs between the AP and the BS. This is the so-called ``hydrogen wall,'' a structure which is much more pronounced in the heliosphere \citep[see][and references therein]{Richardson-etal-2023} and other stars \citep{Edelman-etal-2019}. The small peak around zero results from the fact that there is a slight dip in the hydrogen speed, and thus, the number density increases slightly. The length scales for the interaction between the ions and the neutrals are too large to build a hydrogen wall: At the AP they are of the order of a few au, which is larger than the distance between the AP and the BS. For a larger astrosphere, the interaction length scale is very similar to that for LHS~1140, but the distances between the AP and BS can be much larger, so that a remarkable interaction can take place, building a hydrogen wall which can be observed \citep[for example][]{Edelman-etal-2019}.

The hydrogen flux is so small that it does not noticeably change the astrosphere's structure, but in other astrospheres, like the heliosphere, this can be different.

The neutrals are unaffected by the structures (TS, AP, and BS) produced by the ion fluid. They will mainly pass through these structures unhindered. The effect of the neutrals is quite indirect by charge exchange, e.g., building the hydrogen wall, slowing down the stellar wind plasma in front of the TS due to momentum and energy loading. Lessons we learned are summarized in the book by \citet{Richardson-etal-2023} discussing the astrosphere around our Sun (i.e., the heliosphere).

Because there is almost no variation of the hydrogen flux, the exoplanet will be submerged in it. The inflow into the planetary atmosphere would be $4.8\cdot 10^5$\,\si{cm^{-2} s^{-1}}, or, for the factor 10 larger ISM hydrogen density, equal to $4.8\cdot 10^6$\,\si{cm^{-2} s^{-1}}. 
In the latter case, the astrosphere also shrinks, as is indicated in Table~\ref{tab:dist}. 
This high hydrogen flux can be responsible for building a water world exoplanet \citep{CadieuxEA2023}.

Unfortunately, because the hydrogen flow inside the BS does not very much differ from the interstellar flow, one will not observe a H-$\upalpha$ flux or a Lyman-$\upalpha$ flux along the line of sight (LOS). 

We also want to strongly emphasize the fact that when magnetic fields are involved, one has to model the astrosphere in full 3D because the Parker spiral field of the stellar wind is generally not aligned with the undisturbed ISM 
magnetic field. Also, the ISM magnetic field vector and the velocity vector are not aligned, which leads to an asymmetric astrosphere. The other important fact is that the TS distance can only be determined in the ideal single-fluid HD model. Including a magnetic field in a single-fluid model will change the distances of all structures. Running a multifluid model makes it even worse because, as already discussed for the 2D case in \citep{Scherer-etal-2008}, different hydrogen number densities lead to different TS, AP, and BS positions. 
To demonstrate that, we have changed the hydrogen number density to 1\,\si{cm^{-3}} in the ISM. The result is presented in Table~\ref{tab:dist}, indicated by the row MM1. All other parameters remained the same. As can be seen, the outer shock structure is slightly pressed inward. In the right panel of Figure~\ref{fig:MHDH}, the asymmetry in the $xz$-plane of the proton density can be seen. This feature is very similar to that with the lower interstellar hydrogen density and to the single-fluid MHD case. In these cases, the asymmetry is caused by the non-alignment of the magnetic field and interstellar wind vector. This also demonstrates the need for a 3D multifluid MHD simulation. Moreover, the hydrogen flux needs to be modeled in 3D Cartesian because in a spherical grid, the inner boundary upwind and downwind needs to be connected, for which it is challenging to identify the correct cores in a parallel run. Moreover, inside the inner boundary sphere, the hydrogen flux is affected by charge exchange, for which no analytic function is known.  

\begin{figure*}
    \centering
    \includegraphics[width=\textwidth]{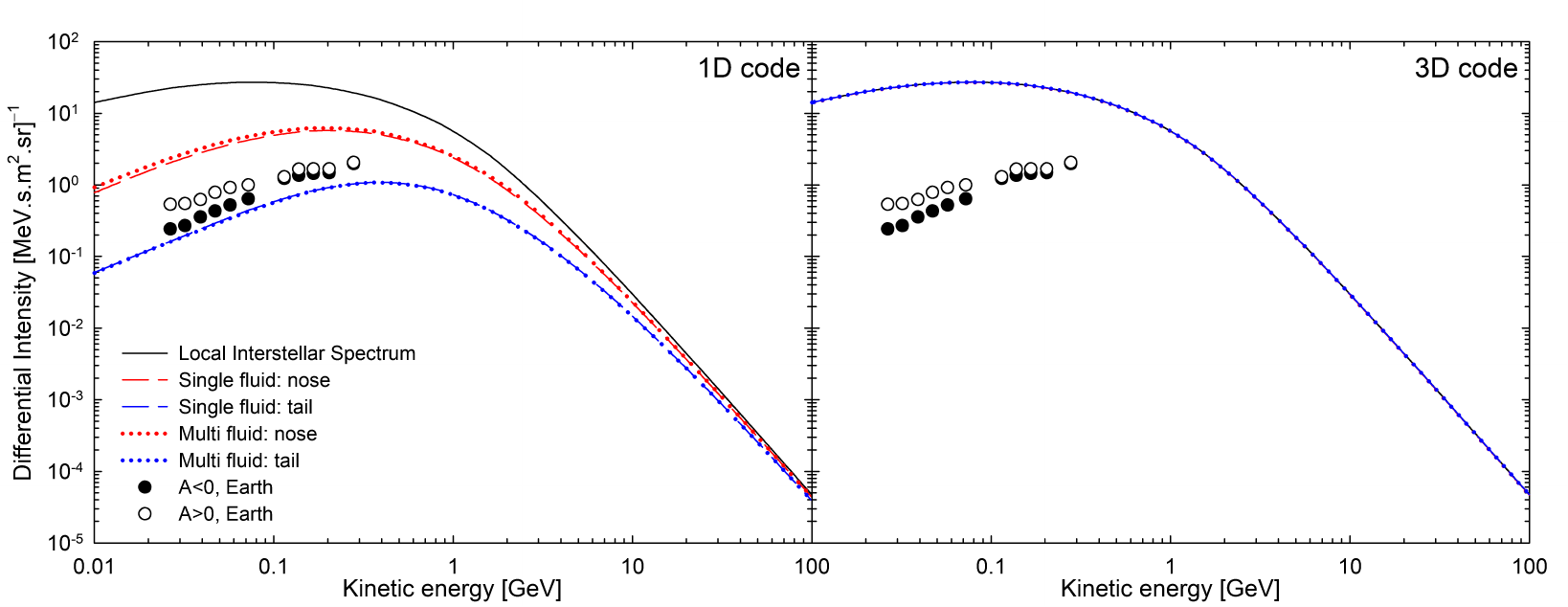}
    \caption{GCR proton differential intensities calculated using a 1D (left panel) and a 3D (right panel) GCR modulation code for the various MHD models under consideration, in the nose and tail directions, as a function of kinetic energy. Also shown are observations of GCR proton differential intensities at Earth reported by \citet{bigmac} to guide the eye. See text for details.}
    \label{fig:gcr}
\end{figure*}
 \section{Modeling GCR transport}\label{sec:GCRs}
Given the consistently small \cite[relative to, say, the heliosphere or other astrospheres, see, e.g.,][]{PogorelovEA17,OpherEA21,HerbstEA20,MesquitaEA21,RodgersEA21,HerbstEA22,KorIz24} size of the astrosphere of LHS~1140, one would expect little to no modulation at LHS~1140~b, with little to no differences in whether one employs results from a single or multifluid MHD model as input to a GCR modulation code. To preliminarily investigate whether this is indeed the case, a 1D stochastic solver based on that introduced by \citet{EngelbrechtDiFelice20} and previously employed in a study of GCR modulation in astrospheres \citep{LightEA22} of the \citet{Parker-1965} transport equation, given by
\begin{align}
    \label{TPE} \nonumber
    \frac{\partial{f}}{\partial{t}} =&\ \frac{1}{r^2}\frac{\partial}{\partial{r}}\left(r^2\kappa_{rr}\frac{\partial{f}}{\partial{r}}\right) + \frac{1}{3\, r^2}\frac{\partial}{\partial{r}}\left(r^2u_\mathrm{sw}\right) \frac{1}{p^2}\frac{\partial}{\partial{p}}\left(p^3f\right) \\
    &- \frac{1}{r^2}\frac{\partial}{\partial{r}}\left(r^2u_\mathrm{sw}f\right),
\end{align}
modeling the influence of diffusion, convection with stellar wind speed $u_\mathrm{sw}$, and adiabatic energy changes on the omnidirectional distribution function $f$ (a function of position and momentum $p$), and hence on the differential intensities of GCRs, is used. The mean free path corresponding to the radial diffusion coefficient $\kappa_{rr}$ is modeled as a function of GCR rigidity following the approach discussed by \citet{LightEA22}, where they are scaled that $\kappa_{rr}\sim1/B$ with $B$ the magnitude of the astrospheric magnetic field, inputs for which being taken along radial spokes in the nose and tail directions of both the single- and multifluid MHD results discussed above. Note that for these calculations, the location of LHS~1140 b is always along the radial spoke in question for ease of comparison. Furthermore, although the GCR local interstellar spectrum (LIS) is expected to vary within the Galaxy \citep{amato}, we employ as a first assumption the LIS of \citet{StraussEA11}. More details about this model can be found in \citet{LightEA22}. 

For comparison, we also compute GCR intensities at LHS~1140~b using a fully 3D solver for the Parker transport equation, now given by
\begin{equation}
\frac{\partial f}{\partial t}
    =
      \nabla \cdot \left( \mathbf{K} \cdot \nabla f \right)
    -  \mathbf{u}_\mathrm{sw} \cdot \nabla f
    +  \frac{1}{3} \left( \nabla \cdot \mathbf{u}_\mathrm{sw} \right) 
            \frac{\partial f} {\partial \ln p},
\label{eq:TPE1}
\end{equation}
which now incorporates a full diffusion tensor $\mathbf{K}$ into the modeling transport perpendicular and parallel to LHS~1140's magnetic field, as well as drift effects. This 3D stochastic modulation code, introduced by \citet{EngelbrechtEA24}, follows exactly the approach outlined in that study: both single and multifluid MHD results are used as inputs for large-scale plasma quantities such as the stellar wind speed, its divergence, and the astrospheric magnetic field magnitude, while small-scale plasma quantities such as magnetic variances and correlation lengths are modeled analytically based on heliospheric observations of the same, but scaled down by the ratio of the magnitude of the astrospheric magnetic field at $1$~au to that of the heliospheric magnetic field. This scaling is motivated by heliospheric observations of turbulence quantities. For example, \citet{BurgerEA22} report that increases in heliospheric magnetic field magnitudes are often accompanied by increases in magnetic variances. More motivations for possible choices of scalings can be found in \citet{HerbstEA22} and \citet{EngelbrechtEA24}. These turbulence quantities are then also used as inputs for the quasilinear/ nonlinear guiding center theory parallel/perpendicular mean free paths \cite[see, e.g.,][]{TS2003,MattEA03,ShalchiEA04,BurgerEA08} used by \citet{EngelbrechtEA24}, the choice of which is motivated again by the usefulness of these expressions in heliospheric GCR modulation studies \cite[e.g][]{MolotoEA18,EW20,EngelbrechtMoloto21}. Note that in this study, the LIS expression of \citet{StraussEA11} is employed in the 3D code also. More technical detail as to the modulation model can be found in \citet{EngelbrechtEA24}.

The left panel of Figure~\ref{fig:gcr} shows GCR proton differential intensities calculated at the location of LHS~1140~b as a function of kinetic energy using the 1D modulation code, alongside observations of GCR intensities at Earth reported by \citet{bigmac} to guide the eye. These observations were taken during periods of positive ($A>0$) and negative ($A<0$) heliospheric magnetic field polarity, where the former denotes the situation where the heliospheric magnetic field points away from the Sun in the northern hemisphere and towards it in the southern hemisphere, the opposite being true for $A<0$. Intensities observed during these epochs in the heliosphere differ due to the influence of drift effects \cite[see, e.g.,][and references therein]{EngelbrechtEA19}. Intensities computed for LHS~1140 for the multi- and single-fluid cases are very similar. It is interesting to note that spectra in the nose and tail directions differ considerably. At LHS~1140, an unexpectedly large amount of modulation can be seen, with intensities in the nose direction being significantly larger than what would be expected at Earth and intensities in the tail direction somewhat smaller, with nose-to-tail differences slightly larger than an order of magnitude at the lowest energy shown. Intensities in the nose direction are lower than those calculated by \citet{HerbstEA20} using a different 1D model and different MHD inputs, highlighting the sensitivity of 1D models to differences in large-scale plasma inputs. This behavior, however, contrasts with the differential intensities at LHS~1140~b yielded by the 3D code, shown in the right panel of Figure~\ref{fig:gcr}. Here, essentially no modulation is seen, with intensities calculated in the nose and tail directions for both the single and multifluid cases being essentially equal to those of the LIS.

\section{Modeling the GCR impact on the atmosphere of LHS~1140~b}
 
In this study, we focus on the impact of GCRs (see Sec.~\ref{sec:GCRs}) on the atmosphere of LHS~1140~b, which -- with its radius of $(1.727 \pm 0.032 R_{\oplus})$ -- is a super-Earth candidate well within the habitable zone of LHS~1140. However, recent spectrally resolved LHS~1140~b observations with the Wide Field Camera~3 (WFC3) on HST suggested that its atmosphere is H$_2$ dominated \citep{EdwardsEA2021}. Utilizing the chemistry-climate code 1D-TERRA, \citet{WunderlichEA2021} presented several possible atmospheric compositions. Here, we focus on a H$_2$-H$_2$O dominated atmosphere with a low amount of CH$_4$ (volume mixing ratio of CH$_4$ of 10$^{-6}$ \% at the planetary surface) that is comparable with the terrestrial pre-industrial era CH$_4$ content \citep{EtheridgeEA1998, WunderlichEA2021}.

Utilizing the Atmospheric Radiation Interaction Simulator \citep[AtRIS, ][]{Banjac-Herbst-Heber-2019}, a GEANT4-based code, we model the propagation of the GCRs in the atmosphere of LHS~1140~b, deriving the GCR-induced atmospheric ionization, a crucial input to chemistry/climate models investigating the impact on observable biosignatures like ozone and methane \citep[e.g.,][]{Herbst-etal-2019b, Scheucher-etal-2020b, Herbst-etal-2024}, and the altitude-dependent absorbed dose rates of a water-based phantom. Therefore, the atmospheric environment was modeled for spherical geometry utilizing a core with a radius of 1.727~$R_{\oplus}$ and a soil composition of 50\% Si, 40\% O, and 10\% Fe. The upper layer of the LHS~1140~b atmosphere was set at $1.534187\cdot 10^{-4}$\,hPa (corresponding to an altitude of 427\,km). To simulate the hadronic and electromagnetic interactions within the H$_2$-H$_2$O dominated atmosphere, we assumed an atmospheric ionization energy ($E_{\mathrm{ion}}$) of 36\,eV \citep[according to][]{SimonWedlundEA2011} and used the Bertini-style cascade for hadrons with energies below 5\,GeV and the Fritiof (FTF) model for particle interactions of mesons, nucleons, and hyperons with energies between 3\,GeV and 100\,TeV (FTFP\_BERT\_HP).

\subsection{The GCR-induced atmospheric ion pair production}
As discussed in, e.g., \citet{Herbst-etal-2019a}, the atmospheric GCR-induced ion-pair production rate $Q_{\mathrm{GCR}}$ as a function of the exoplanetary magnetic field and atmospheric pressure $x$ is given by
\begin{equation}
    Q_\mathrm{GCR}(E_\mathrm{C}, x) = \sum_i \int_{E_\mathrm{C}}^{\infty} \alpha \, J_i(E) \frac{1}{E_{\mathrm{ion}}} \,\frac{\Delta E_i}{\Delta x} \, \mathrm{d}E \,.
\end{equation}
Here, $E_C$ represents the so-called cutoff energy, the energy a particle must have in order to enter a planetary magnetic field at a particular location and altitude (i.e., pressure). Since -- so far -- it is not clear whether rocky exoplanets around M~stars have magnetic fields such as those known from solar system bodies, only a very weak tilted dipole field (i.e., with a cutoff energy of $E_\mathrm{C}$ = 1\,MeV) has been assumed in this study. Further, $J_i(E)$ represents the differential GCR energy spectrum of type $i$ (i.e., protons, He, etc.)  as discussed in Sec.~\ref{sec:GCRs}, $\alpha = 2 \pi \int \cos({\theta}) \sin({\theta}) \, \mathrm{d} \theta$ is the geometrical normalization factor with $\theta$ being the particle's angle of incidence, $\Delta E_i$/$\Delta x$ the modeled pressure-dependent mean specific energy loss of the GCRs within the exoplanetary atmosphere, and \mbox{$E_{\mathrm{ion}}$ = 36\,eV}.

\begin{figure}[t!]
    \centering
    \includegraphics[width=\columnwidth]{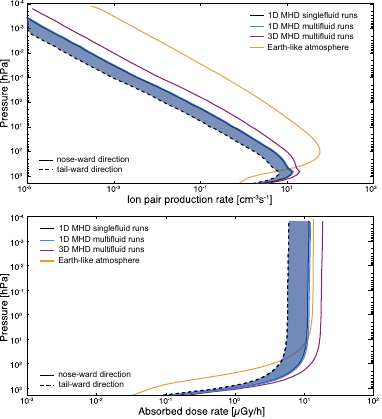}
    \caption{Upper panel: GCR-induced atmospheric ionization of LHS~1140~b based on the 1D single- and multifluid runs (in black and blue, respectively) in noseward (solid lines) and tailward (dashed lines) direction. The results utilizing the 3D GCR fluxes are shown in purple. In comparison, the results for an Earth-like atmosphere of TRAPPIST-1e  \citep[e.g., ][]{Herbst-etal-2024} are highlighted in orange. Lower panel: GCR-induced absorbed dose rates. Colors and line styles according to the upper panel.}
    \label{fig:ippr}
\end{figure}

The upper panel of Figure~\ref{fig:ippr} shows the GCR-induced ion pair production rates of our four investigated scenarios. Due to the tail/nose differences in the 1D model runs of the single-fluid (in black) and the multifluid (in blue) astrospheric input, strong modulation of the induced atmospheric ionization by a factor of almost two at the surface (with a pressure of 2584.47\,hPa) up to a factor of four at a pressure of $1.53\cdot 10^{-4}$\,hPa (at 427 km) is visible (shaded areas). However, utilizing the 3D model results induces much higher ionization rates (purple line) throughout the entire atmosphere. Even at the surface of LHS~1140~b, this leads to an ionization increase of 25\% compared to the 1D multifluid results. A comparison with the proton-induced values in the Earth-like wet and alive CO$_2$ rich atmosphere of TRAPPIST-1e discussed in \citet{Herbst-etal-2024} (comparable surface pressure, orange line) indicates that the impact of GCRs at the surface of LHS~1140~b is more severe by showing ionization rates that are four times higher. 
\subsection{The GCR-induced atmospheric radiation exposure}
Following the approach discussed in, e.g., \citet{HerbstEA2020}, we further modeled the atmospheric radiation exposure in the H$_2$-H$_2$O dominated atmosphere of LHS~1140~b. Utilizing a water-based phantom \citep[i.e., the ICRU phantom, mimicking the human body, see][]{McNair1981}, the mean absorbed dose rates can be derived by
\begin{equation}
 \mean{D}_j(E_i, r) = \left(\frac{E_\mathrm{d}}{E_i}\right)_j \, \frac{E_i}{(4/3) \pi \, r_{\mathrm{p}}^3 \, \rho_{\mathrm{p}}},
\end{equation}

where $E_\mathrm{d}$ represents the average ionization energy of a particle $j$ within the phantom with density $\rho_\mathrm{p}$ and mass $m_\mathrm{p}$ and $E_i$ is the kinetic energy of the primary particle. The altitude-dependent absorbed dose rates presented in the lower panel of Figure~\ref{fig:ippr} are derived by a convolution with the GCR spectra and summing up over all energy bins and particles. 

As shown, the GCR-induced absorbed dose rates vary between \mbox{0.092 $\upmu$Gy/h} (based on the 1D single-fluid results) and \mbox{0.2 $\upmu$Gy/h} (based on the 3D multifluid results) at the surface. While the latter is comparable with the absorbed dose rates at Earth's surface of \mbox{0.5 $\upmu$Gy/h}, albeit with a much lower surface pressure of 1033\,hPa \citep[e.g.,][]{HerbstEA2020}, a comparison with the N$_2$-O$_2$ dominated Earth-like scenario of TRAPPIST-1e (orange line) shows an order of magnitude higher dose rate at the surface of LHS~1140~b.

\section{Discussion}\label{sec:discussion}

\subsection{On the influence of the stellar magnetic field strength}
In this study, we assumed a stellar magnetic field strength of 1 \si{G}. According to \citet{2024MNRAS.527.4330L}, the large-scale magnetic field of selected slowly rotating M stars seems to vary within 20 to 200 G. It should be noted, however, that the reported magnetic tilt angles of the four stars comparable with LHS\,1140 in terms of rotation period indicate that at least two of them (i.e., GJ 905 and GJ 1151 with changes of 61$^\circ$ and 87$^\circ$, respectively) can be construed to be undergoing a transition between stellar cycles. Although such tilt angle variations are in agreement with the solar wind magnetic field during a Schwabe cycle, the reported magnetic field strength changes are in the order of 64 \si{G} (GJ~905) and 39 \si{G} (GJ 1151). Because of the high variability of the reported magnetic fields, our assumption of an averaged magnetic field of 1 \si{G} for LHS\,1140 thus might be equally appropriate. 

\begin{figure}[]
    \centering
    \includegraphics[width=\columnwidth]{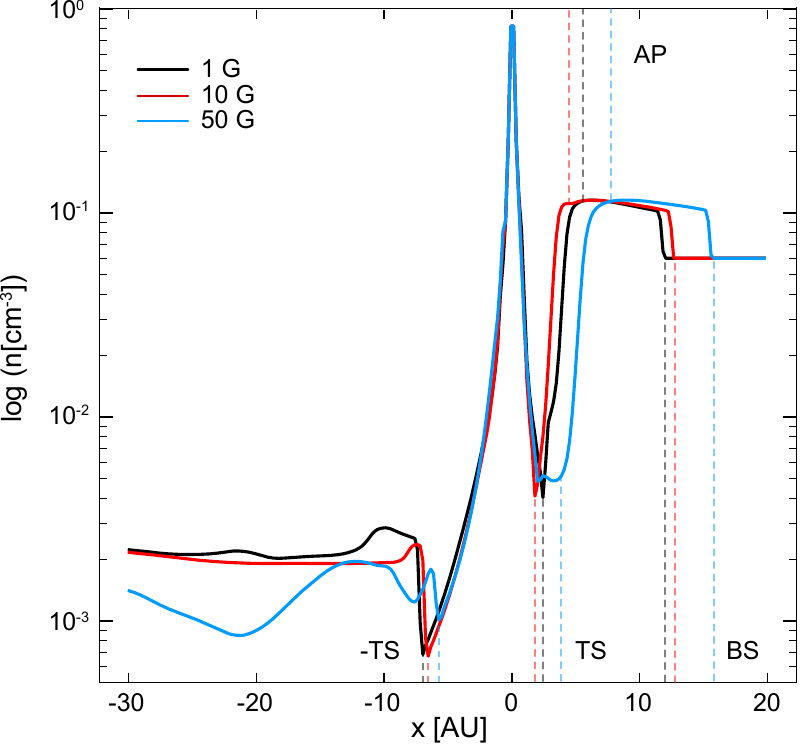}
    \caption{Visualization of the TS, AP, and BS distance changes due to variable stellar magnetic field strengths (1 \si{G} in black, 10 \si{G} in red, and 50 \si{G} in blue). The number density in the rest of the star is shown.}
    \label{fig:Bvar}
\end{figure}

Nevertheless, to study the impact of a stronger large-scale magnetic field on the size of the astrospheric structure of LHS\,1140, we repeated our model efforts, assuming a 10 \si{G} and 50 \si{G} large-scale magnetic field strength. The results are displayed in Fig.~\ref{fig:Bvar}. Although there is a slight shift towards larger TS, AP, and BS distances with increasing stellar magnetic field strength (see also Table~\ref{tab:Bvar} for further information), the modulation of GCRs within, however, is unaffected by these changes (not shown here).

Because we always ensure that the Alfv\'en surface, i.e., \ the surface where the Alfv\'en Mach number $M_\mathrm{A} = 1$, is inside our integration area, the ram pressure dominates over the magnetic field pressure:
\begin{align}\label{eq:MA}
   1> \dfrac{\rho u^{2}/2}{B^{2}/(8\pi)} = M_{A}^{2} .
\end{align}
If $M_\mathrm{A}<1$, assuming a Parker spiral given by
\begin{align}
    \pm \vec{B}_{\mathrm{sw},\star} = B_0 \frac{r_0^2}{r^2} \left(\vec{e}_{r} -\frac{\Omega_\star r}{r_0 u}\sin\vartheta \, \vec{e}_{\vartheta}\right),
\end{align}
at the inner source is then not possible (note that $r_0$ and $B_0$ denote the reference distance and magnetic field respectively, $\Omega_\star$ the stellar rotation period, and $\vartheta$ the co-latitude). Consequently, the magnetic pressure does not play a significant role. The case with $B_\star=50\si{G}$ has a magnetic field that dominates the inner astrosheath. This case needs a more detailed study. 

To get an idea of the TS distance, we assume a 1\,kG dipole magnetic field at the surface of the star ($r_0=R_\star$), and set this magnetic field pressure equal to the ram pressure of the interstellar medium:
\begin{align}
    \frac{B^2(r)}{8\pi} = \frac{1}{2} \rho_{\mathrm{ism}} v_{\mathrm{ism}}^2
\end{align}
For simplicity, we consider only the $r^{-3}$-dependency of the dipole field strength, with ISM parameters given in Appendix~\ref{app:params}, we get for the TS distance
\begin{align}
    r_{TS} &= \left(\frac{4\pi B(r_0)^2}{\rho_{\mathrm{ism}} v_{\mathrm{ism}}^2}\right)^{1/3} r_0\approx 308 \,\si{au} \ .
\end{align}
This estimate is much too large, because the thermal pressure and ram pressure will have some influence, thus a complete new model setup is needed. 

On the other hand, we do not expect such high magnetic fields because LHS1140 is a very slow rotator; thus, the magnetic dynamo will not be very effective. This justifies our choice of the magnetic field to have values below 50\,G. 

\begin{table}[!b]
\caption{\label{tab:Bvar} TS, AP, and BS distances}
\centering
    \begin{tabular}{|c|c|c|c|c|c|}
    \hline
         Model & TS & AP & BS &Analytic \\
         \hline
        $B_\star$ = 1\,\si{G} &2.2   &3.7 &11.4 & 3.3\\ 
        $B_\star$ =10\,\si{G} &1.9   &3.0 & 12.3 & 3.3\\
        $B_\star$ = 50\,\si{G}&2.2   &3.7 & 15.5 & 3.3\\
        \hline
       $\dot{M}_\star = 1.2\cdot10^{-15}\, {\dot{M}_\odot}$ &\mr{10.4}  &\mr{23}  & \mr{45}  & \mr{16.2}\\
        $v_\mathrm{sw}$ = 250  \,\si{km/s} &&&&\\
       \hline
       $\dot{M}_\star = 1.2\cdot10^{-15} {\dot{M}_\odot}$&   \mr{20} & \mr{46}   &\mr{96} &  \mr{32.3}\\
       $v_\mathrm{sw}$ = 1000\,\si{km/s} &&&&\\
       \hline
       $\dot{M}_\star = 2.6\cdot10^{-16} {\dot{M}_\odot}$&   \mr{7.8} & \mr{15}   &\mr{37} &  \mr{10.9}\\
       $v_\mathrm{sw}$ = 430\,\si{km/s} &&&&\\
 \hline
    \end{tabular}
    \tablefoot{TS, AP, and BS distances for the model runs discussed in Sec.~\ref{sec:discussion}. Note that in the cases where only the magnetic field strength was changed (rows 1 to 3), the analytic solutions are always identical because the magnetic field is neglected in Eq.~(\ref{eq:ts_dist}). Rows 4 to 6 show the results based on assuming a magnetic field strength of $B_\star$ = 1\si{G} while changing the mass-loss rate (in rows 4 and 5 set to $1.2\cdot10^{-15}\, M_\odot/\si{yr}$) and the stellar wind speed ($v_\mathrm{sw}$ = 250 km/s, row 4 and $v_\mathrm{sw}$ = 1000 km/s, row 5). Row 6 shows the results for $2\cdot10^{-16}\, M_\odot/\si{yr}$ and $v_\mathrm{sw}$ = 250 km/s.}

\end{table}

\subsection{On the influence of the stellar mass loss rate}

In this study, we assumed a stellar mass-loss rate of 5$\cdot$10$^{-17}$ $M_\odot$ yr$^{-1}$, which was based on the fact that LHS\,1140 is a slow rotator showing little to no stellar activity. The assumed value is also still above the lowest mass loss rate of an M4.5 star reported in \citet{10.1093/mnras/stx1543} (i.e., $> 2.2$ $\cdot 10^{-17} M_\odot$ yr$^{-1}$). According to \citet{2021ApJ...915...37W}, who derived the mass loss rates of 17 M~stars, GJ 406 (M4.5, rotation period $\sim$ 40 days) is listed with the lowest mass loss rate value of 0.059 $\dot{M_\odot}$ \citep[which is also in agreement with the value reported by][]{10.1093/mnras/stx1543}. We note that \citet{2021ApJ...915...37W} utilized stellar X-ray fluxes to derive the stellar mass loss rates. \citet{Cohen-2011} showed that the solar X-ray flux varies by an order of magnitude over a solar cycle and found that the solar mass loss varied by a factor of two to five during the observations between 1996 and 2006. Thus, the newest data provided by \citet{2021ApJ...915...37W} should be corrected for such effects and -- for now -- can only provide a sophisticated estimate of the input data. Moreover, \cite{Wood-etal-2021} also stated the necessity to include magnetic fields in the modeling.

\begin{figure}[t!]
    \centering
    \includegraphics[width=\columnwidth]{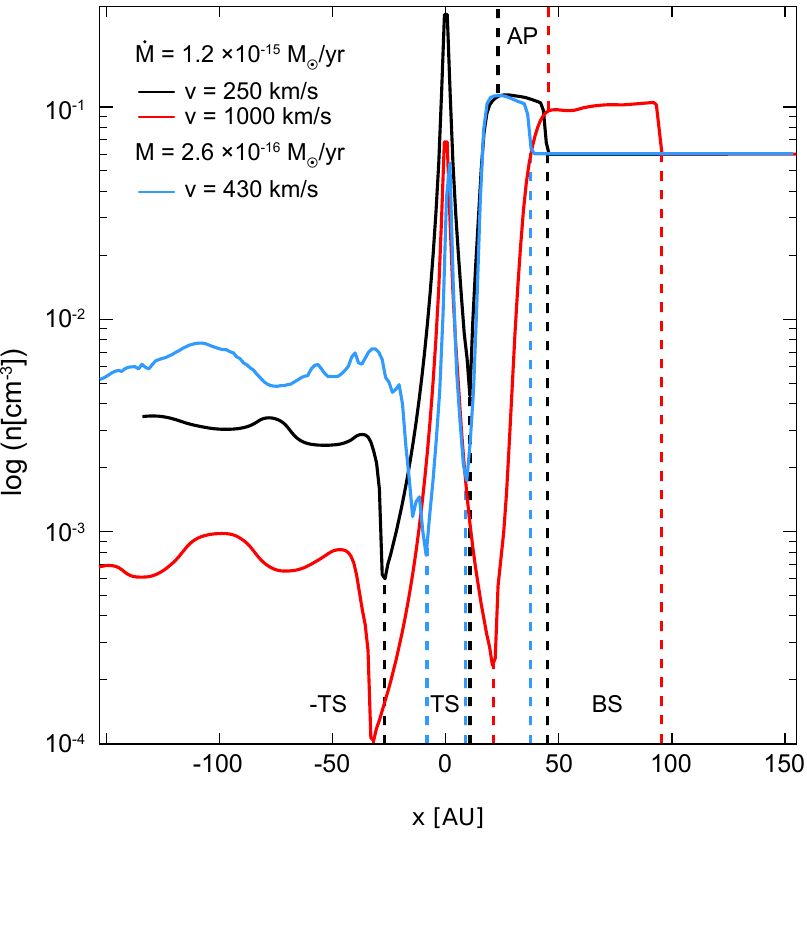}
    \caption{TS, AP, and BS distance changes caused by different stellar wind speeds: $ v_\mathrm{sw}=250\, \si{km/s}$ (black line) and $ v_\mathrm{sw}=1000\, \si{km/s}$ (red line). In both runs a stellar mass loss rate of $\dot{M} = 1.2\cdot10^{-15}\, M_\odot/\si{yr}$ was assumed. The blue line shows the distances for $ v_\mathrm{sw}=430\, \si{km/s}$ and $\dot{M} = 2.6\cdot10^{-16}\, M_\odot/\si{yr}$.}
    \label{fig:Mvar}
\end{figure}

To study the impact of a higher mass loss rate on the shape and size of the astrosphere of LHS~1140, model runs were performed utilizing a stellar mass loss rate of 5$\cdot 10^{-15}$ $M_\odot$ yr$^{-1}$ (i.e., $\dot{M} = 0.059 \dot{M}_{\odot}$). We also increased the wind speed to $v_\mathrm{sw}=1000\,\si{km/s}$ in a second run. We also show the TS, AP, and BS distances for the stellar wind data derived in Appendix~A. A comparison with the original run is shown in Fig.~\ref{fig:Mvar}, where it is clear that higher mass loss rates and stellar wind speeds can lead to a significantly larger astrosphere. This behavior is also indicated by Eq.~(\ref{eq:ts_dist}), which also suggests that a higher mass loss rate and stellar wind speeds lead to a larger TS distance, which still has to be determined by a 3D multifluid MHD model. Nevertheless, all discussed cases show shorter distances than those for the heliosphere.

\subsection{On the uncertainties of the LISM parameters}

The ISM parameters are always a guess. For nearby stars, we can use the number density, temperature, and magnetic field for the heliosphere. For the ISM velocity, we always use the radial speed and proper motion as described in \cite{Wood-etal-2021}. The direction and velocity of the ISM are not known and might vary dramatically, as can be seen in the O-star bow shock sample \citep{Peri-etal-2012,Peri-etal-2015}. Therefore, we only use the radial speed and the proper motion to determine the relative speed between the star and the ISM (i.e., the ISM wind speed). This assumes that the ISM is at rest relative to the 
The hydrodynamic multifluid models have the advantage that only a half-plane needs to be modeled since they are axi-symmetric. By including magnetic fields, the stellar wind and the ISM magnetic field break this symmetry, and a fully 3D model is needed. Moreover, because of this symmetry break, the line-of-sight integration will generally lead to different neutral H-regimes compared to the HD one. In addition, the magnetic pressure adds to the total pressure in the momentum and energy equation and thus reduces the astrospheric distances to the TS, AP, and BS. Because of said asymmetry, the astrosphere must be rotated into the correct position and placed into a skymap. This procedure is discussed in \citep{Baalmann-etal-2020} for a spherical case. 

\citet{Wood-etal-2021} are using solar wind data for the stellar winds, which is unreliable for M~stars with much higher wind speeds. Moreover, these observations are only a snapshot because stars have cycles \citep[for example
][and references
therein]{Althukari-Tsiklauri-2023,Bondar-2019,Jeffers-etal-2023}, and
\citep[especially for exoplanets][]{Obridko-etal-2022}like the solar cycle. During such a cycle, the ram pressure can vary, and thus, the observed fluxes can change. For example, the X-ray flux of the Sun changes by a factor of ten, while its mass loss rate can vary by a factor of 2--5 \citep{Cohen-2011}.  

The original determination of the X-ray flux \citep{Wood-Linsky-1998} is based on the ram pressure of an ideal HD model.  The neutral hydrogen flow causes a momentum and energy loading of the plasma, and the stellar wind in front of the TS is decelerated and heated. This will affect the interpretation of the X-flux determination. Things become even more difficult when a magnetic field is involved because the magnetic field will play a role (via the magnetic field pressure and the magnetic stress tensor). This also applies when the Lyman-$\alpha$ flux is used instead of the X-flux.
\citet{Wood-etal-2021} also states that a full 3D multifluid simulation is needed to improve their results.

\section{Summary and Conclusion}
We have shown that a fully 3D multifluid MHD simulation is necessary to model the astrospheres and the GCR flux because the magnetic fields break the symmetry. In particular, we have shown that single-fluid (M)HD or multifluid HD simulations lead to larger distances for the shock structures and, moreover, that Eq.~(\ref{eq:ts_dist}) is only valid in an ideal single-fluid HD case.  We have shown that the neutral hydrogen flux does not change inside the astrosphere, and thus, the exoplanet is submerged in the ISM hydrogen flux.

As one important result, we have shown that for different input parameters (like stellar magnetic field, mass loss rate, and speed) the analytic estimate of Eq.~(\ref{eq:ts_dist}) does not hold (see Table~\ref{tab:Bvar}). Because almost all cool stars have an intrinsic magnetic field besides the ISM magnetic field, a 3D multifluid model  is needed  to determine the modulation volumes for the GCRs.

As expected for such a small astrosphere, the 3D GCR modulation code of \citet{EngelbrechtEA24} yields essentially unmodulated GCR proton intensities at LHS~1140~b, regardless of the planet's azimuthal placement in the model,  when run utilizing the results of the 3D MHD approach outlined in Section~\ref{sec:2}. This contrasts strongly with what is seen when the 1D GCR modulation code of \citet{LightEA22} is employed  using the same MHD inputs: computed intensities display significant levels of modulation and differ strongly when the planet is located at an azimuth corresponding to the nose versus the tail direction. These differences are due to the fact that the 3D code includes transport effects that cannot be considered in a 1D model. A pertinent example of this are the differences in diffusion parallel and perpendicular to the astrospheric magnetic field. LHS~1140 has a long rotation period of $\sim131$ days \cite[e.g.][]{Lillo-BoxEA2020} relative to, say, the Sun, which leads to a strongly underwound magnetic field, which can facilitate the inward transport of GCRs \cite[see the discussion by][]{EngelbrechtEA24}. Azimuthal variations in GCR intensities yielded by the 1D model are also not apparent in results from the 3D model, as both azimuthal and latitudinal transport of GCRs is possible in the latter code. This study highlights the dangers implicit in using a lower-dimensional GCR modulation model, in that the results of such an approach may be unrealistically small, even for a tiny astrosphere such as that of LHS~1140, or imply potentially unphysical spatial variations which could unduly influence the results as to, for example, atmospheric ion pair production calculated therefrom. It should be noted that GCR spectra from the 3D GCR code could be expected to be more modulated for the larger astropsheres computed for higher stellar wind speeds and mass loss rates (see Fig.~\ref{fig:Mvar}) but would not be significantly lower, due to the slow rotation of LHS\,1140. Nevertheless, this highlights the importance and necessity of improved observational estimates of input parameters for the MHD modeling approach discussed here.

When the atmospheric ionization results utilizing the 1D and 3D GCR transport are compared, the latter is shown to vary between 25\% at the surface of LHS~1140~b and 220\% in the upper atmosphere. Since atmospheric chemistry processes, and with those, the derivation of transmission spectra features and information on biosignatures strongly depend on atmospheric exoplanetary ionization \citep[e.g.,][]{Herbst-etal-2019b, Herbst-etal-2024}, our results show that for inactive cool stars such as LHS~1140 reliable GCR-induced background radiation information is mandatory to interpret, e.g., upcoming JWST data.

\begin{acknowledgements} 
The authors gratefully acknowledge support from the German Research Foundation (Deutsche Forschungsgemeinschaft, DFG) under grant \textit{HE 8392/2-1} (project 508335258) to initialize this international collaboration and thank the Center for High Performance Computations (CHPC) in Cape Town for partially funding it under the project \textit{ASTRO1277}. KS acknowledges the support of the DFG grant \textit{SCHE 334/16-1} (project 491027218). JK acknowledges DFG support through the Collaborative Research Center (Sonderforschungsbereich, SFB) 1491 ``Cosmic Interacting Matters -- From Source to Signal.'' KH acknowledges the support of the DFG priority program SPP~1992 ``Exploring the Diversity of Extrasolar Planets (\textit{HE 8392/1-1}).'' This work is based on the research supported partly by the National Research Foundation of South Africa (NRF grant number 137793). Opinions expressed and conclusions arrived at are those of the authors and are not necessarily to be attributed to the NRF. The authors gratefully acknowledge the Gauss Centre for Supercomputing e.V. (\url{www.gauss-centre.eu}) for funding this project by providing computing time through the John von Neumann Institute for Computing (NIC) on the GCS Supercomputer JUWELS at Jülich Supercomputing Centre (JSC).
\end{acknowledgements}
 \bibliographystyle{aa}

\begin{appendix}
\section {Stellar wind and ISM parameters and their implementation in the Cronos Code}
\label{app:params}

We use the approach described in
\citet{Modi-etal-2023} (hereafter denoted as MEV) to determine the stellar wind speed. First, we take the age of LHS\,1140 from Table~5 in \citet{Engle-Guinan-2023} as $\tau = 7.84$\,Gyr. We use the logarithm of the X-ray luminosity 
\begin{align} \nonumber
 \log \left( L_\mathrm{X} [\si{erg\, s^{-1}}] \right) &= -1.4214 \log\left(\tau [\si{Gyr}] \right) +27.826 \\
 &= 26.56
\end{align}
according to Eq.~(4) in MEV and the resulting X-ray flux
\begin{align}
  F_\mathrm{X} = \frac{L_\mathrm{X}}{4\pi R_{\star}^{2}} = 1.34 \cdot 10^{5}\,\si{erg\, s^{-1} cm^{-2}} \ ,
\end{align}
from which the temperature can be calculated as
\begin{align}
  T = 0.11\,\si{MK} \left(F_{X} [\si{erg\, s^{-1} cm^{-2}}]\right)^{0.26}  = 2.4 \,\si{MK}
\end{align}
according to Eq.~(4) in \citet{Johnstone_Guedel:2015}.
Thus, the thermal speed yields 
\begin{align}
    v_c = \sqrt{\frac{\gamma k_\mathrm{B}\, T}{m_\mathrm{p}}} = 182\, \si{km/s} \ ,
\end{align}
where $\gamma=5/3$ is the polytropic index for a proton plasma, $k_\mathrm{B}$ the Boltzmann constant, and $m_\mathrm{p}$ the proton mass.

We can now apply the isothermal Parker wind model to estimate the stellar wind speed. 
The critical
radius for the Parker stellar wind model is
\begin{align}
  r_\mathrm{c} = \frac{G M_{\star}}{2 \, v_\mathrm{c}^{2}}
= 0.0024\,\si{au} = 2.4\,R_\star. 
\end{align}
We assume that the stellar wind has reached its final speed at $r=10\,R_\star$ so that
\begin{align}
  v_\mathrm{sw} = 2 v_\mathrm{c} \sqrt{\ln(r/r_{c})} \approx 430 \, \si{km/s}. 
\end{align}
There are other, more advanced stellar wind models, like the AWSoM model
used by \citet{Chebly-etal-2023}, but these models require detailed
observational input, which is not available for LHS\,1140. 

We take the stellar wind density $\rho =2.75\cdot 10^{-22}\,\si{g \, cm^{-3}}$ at LHS1140~b ($r_\mathrm{p} = 0.0270\,\si{au}$) after 7.84\,\si{Gyr} according to MEV and get
\begin{align}
    \dot{M}_\star = 4 \pi \rho v_\mathrm{sw} r_\mathrm{p}^2 = 2.6 \cdot 10^{-16} M_\odot/\si{yr} \approx 0.01 {\dot{M}_\odot}
\end{align}
with $v_\mathrm{sw} = 430\,\si{km/s}$. 

Given that the solar mass loss rate varies by a factor of 2 to 5, we set the
mass loss rate of LHS\,1140 to $5\cdot 10^{-17} M_{\odot}/\si{yr}=
0.0025 \dot{M}_{\odot}$ and $v_\mathrm{sw} = 250\,\si{km/s}$ as lower limit. These models already existed before we were aware of the above values and are sufficient for the comparison of the different HD and MHD models shown in
Figs.~\ref{fig:HD} and~\ref{fig:MHDp}. The more realistic results derived above are shown in Fig.~\ref{fig:Mvar} together with other model parameters. Because the contour plot looks very similar (except for the scaling), we did not show them.

As described above, our model uses the Parker spiral magnetic field, which gives a frozen-in magnetic field outside the Alfv\'en radius. To apply this, the Alfv\'en radius must be smaller than the termination shock distance to avoid waves moving backward into the boundary. Beyond the Alfv\'en radius, the flow is super fast-magnetosonic and supersonic. Therefore, we have chosen the initial magnetic field to $B=1\,\si{G}$ and discussed also results based on magnetic field strengths of up to 50 \si{G}. Note, however, that for our setup, no stronger magnetic fields can be assumed because otherwise, the Alfven radius is larger than the TS distance and, with that, reaches into the shock structure. Investigating such a scenario will be the subject of future work.

We took the thermal pressure of the solar wind from the near-Earth satellites that have measured the solar wind since 1955 (IMP8, ACE, SoHo, WIND, etc.), which is usually around 70{,}000\,K but varies strongly during solar cycles (between 50{,}000 and 150{,}000\,K according to \cite{Shi-etal-2023}), and between 3000\,K and $\sim 2$\,MK according to the WIND data taken between 2000 and 2024 available here\footnote{\url{https://omniweb.gsfc.nasa.gov/form/sc_merge_min1.html}}. The kinetic electron temperature, as discussed in \cite{Wilson-etal-2018} and \cite{Scherer-etal-2022}, is different from the large-scale MHD data. We scaled the surface temperatures of the Sun and LHS\,1140 to an average solar wind proton temperature at 1\,\si{au} and got 46~000\,K. These temperatures do not play a role in the dynamics because the ram pressure is much larger. At the termination shock, where the thermal pressure is small, about three-quarters of the ram pressure is converted into thermal pressure beyond the shock. Thus, one has only to guarantee that the sound speed at the Alfv\'en surface is larger than one, and hence, the thermal pressure is small compared to the ram pressure.

 We assume that the stellar wind parameters are known at a stellar distance $r_0=1\,\si{au}$ and the ISM parameters at infinity. We assume for the stellar wind number density and thermal pressure a spherical expansion (i.e., $\rho(r) \propto r^{-2}$ and $p(r) \propto r^{-10/3}$) for $r<1\ \si{au}$ until the Alfv\'en radius is reached. That is not the case for the resolution adopted here, as the Cartesian grid size for the cell centered at the origin is always larger than the Alfv\'en radius. For $r>1$, the stellar wind is expanded spherically until it reaches $r_1 := r_\mathrm{TS}/2$ as given by Eq.~(\ref{eq:ts_dist}) while cells at distances larger than $r_2:= 2\, r_\mathrm{TS}$ are filled with the ISM values. Between these two radii, the stellar wind parameters linearly approach the interstellar parameters (see below). 
The magnetic field at $r_0$ is assumed to behave like a Parker spiral, giving the values for all other cells up to $r_1$. 
To guarantee that the magnetic field is divergence-free, the magnetic fields for the stellar wind and the ISM are represented through their respective vector potentials:
\begin{align}
    \vec{A}_\mathrm{sw} &= B_\mathrm{sw} \left(
       -  \frac{2\pi}{P_\mathrm{rot} \, v_\mathrm{sw}} \, |\cos\vartheta| \ \vec{e}_r
       + \frac{1-\cos\vartheta}{r \sin\vartheta} \ \vec{e}_\varphi
    \right) \\
    \label{eq:A_ism}
    \vec{A}_\mathrm{ism} &= \frac{1}{2}
    \begin{pmatrix}
    B_{y,\mathrm{ism}} \, z - B_{z,\mathrm{ism}} \, y  \\
    B_{z,\mathrm{ism}} \, x - B_{x,\mathrm{ism}} \, z  \\ 
    B_{x,\mathrm{ism}} \, y - B_{y,\mathrm{ism}} \, x  
    \end{pmatrix} ,
\end{align}
with 
\begin{align}
    B_{x,\mathrm{ism}}  &= B_\mathrm{ism} \sin(\vartheta_{B_\mathrm{ism}})  \cos(\varphi_{B,\mathrm{ism}})\\
    B_{y, \mathrm{ism}} &= B_\mathrm{ism} \sin(\vartheta_{B_\mathrm{ism}})  \sin(\varphi_{B,\mathrm{ism}})\\
    B_{z, \mathrm{ism}} &= B_\mathrm{ism} \cos(\vartheta_{B_\mathrm{ism}})
\end{align}
the Cartesian components of $\vec{B}_\mathrm{ism}$. The total magnetic field is then obtained via
\begin{equation}
    \vec{B} = \nabla \times \left[(1-f) \vec{A}_\mathrm{sw} + f \vec{A}_\mathrm{ism} \right] ,
\end{equation}
where $f(r)$ is zero inside $r_1$, unity outside $r_2$, and linearly increasing for $r_1 \le r \le r_2$.
The velocity is kept constant in the stellar wind and ISM, and between $r_1$ and $r_2$  a linear transition, as described above, is adopted.

Table~\ref{tab:2} summarizes known properties of LHS~1140 and parameters used for the simulations in this work. The data have been updated from \citet{HerbstEA20} using the latest results for the radial velocity and proper motion from the \href{https://simbad.cds.unistra.fr/simbad/sim-basic?Ident=LHS1140}{Simbad data base}, from which the ISM speed is calculated.

\begin{table}[h!]
\caption{\label{tab:2} 
Model parameter}
\begin{tabular}{l|r|l|l}
Parameter & LHS~1140 & Sun &Comment \\
\hline
type & M4.5 &G2V & (a) \\
$T_{\mathrm{eff}}$ & 3220\,K &5772\,K & (b) \\
$P_{\mathrm{rot}}$ & 131\,d & 25\,d (equator) & (b) \\
$L_{\star}/L_{\odot}$ & 4.4 $\cdot 10^{-3}$& 1& (c) \\
$R_{\star}/R_{\odot}$ & 0.212 & 1&(b) \\
$M_{\star}/M_{\odot}$ & 0.179 &1 &(b) \\
$B_{\star}$ &  1\,G& 1\,kG& (d,i) \\
$\dot{M}_{\star}$ & $5 \cdot 10^{-17} M_{\odot}$ yr$^{-1}$ & $2 \cdot 10^{-14} M_{\odot}$ yr$^{-1}$ & (e) \\
$d_{\star}$ &14.98\, pc & 1\, au&\\
age & 7.84\, Gyr & 4.6\, Gyr & (f)\\
\hline
$T_{\mathrm{sw}}$ & $46 \cdot 10^3$\,K & $70 \cdot 10^3$\,K & (g) \\
$u_{\mathrm{sw}}$ & $250$\,km s$^{-1}$ & 400\, km s$^{-1}$ & (e)\\ 
$n_{\mathrm{sw}}$ & 0.04\,cm$^{-3}$ & 4\,cm$^{-3}$ &(e) \\
$B_{\mathrm{sw}}$ & 0.33\,nT &  4\, nT& (g) \\
\hline
$T_\mathrm{ism, p}$ & 12000\,K & 6370\,K&(g) \\
$u_\mathrm{ism, p}$ & 48\,km s$^{-1}$ & 26\, km s$^{-1}$ & (g) \\
$n_\mathrm{ism, p}$ & 0.06\,cm$^{-3}$ & 0.06\,cm$^{-3}$ & (g) \\
$B_\mathrm{ism}$ & 0.3\,nT & 0.3\,nT & (g) \\
$\varphi_{B, \mathrm{ism}}$ & $150^\circ$ & $150^\circ$ & (g) \\
$\vartheta_{B, \mathrm{ism}}$ & $30^\circ$ &  $30^\circ$&  (g) \\
$T_\mathrm{ism, H}$ & 12000\,K & 6370\,K&f(g) \\
$u_\mathrm{ism, H}$ &  48\,km s$^{-1}$ & 26\, km $s^{-1}$ & (g) \\
$n_\mathrm{ism, H}$ &  0.1\,cm$^{-3}$ & 0.1\,cm$^{-3}$&(g) \\ \hline
\end{tabular}
\tablefoot{ 
Stellar properties of LHS~1140 (first block), the corresponding stellar wind properties at 1\,au
  (second block), and the assumed ISM parameters (third block). Solar values are listed for comparison. The direction of the inflow vector is always along the $x$-axis. Comments: (a) = Data taken from the \href{https://simbad.cds.unistra.fr/simbad/sim-basic?Ident=LHS1140\&submit=SIMBAD+search}{Simbad data base}, 
(b) = taken from \citet{Benedict-etal-1998},
(c) = data calculated via the Stefan-Boltzmann law $L_{\star} \propto R_{\star}^2 T_{\star}^4$, 
(d) = sophisticated guess, see text, 
(e) = after \citet{Modi-etal-2023},
(f) = after \citet{Engle-Guinan-2023}
(g) = sophisticated estimation based on heliospheric parameters.
(i) the solar magnetic field strength seen from 1\,pc (private communication with P.J. Steyn).
}

\end{table}

\end{appendix}

\end{document}